\documentclass[aps,eqsecnum,showpacs,twocolumn,floatfix]{revtex4}
\usepackage{graphicx}
\usepackage{amsmath}
\usepackage{amsfonts}
\usepackage{amssymb}
\usepackage{bm}
\begin{document}

\title{Detecting groups of similar components in complex networks}
\author{Jiao Wang$^{1,3}$ and C.-H. Lai$^{2,3}$}
\affiliation{$^{1}$Temasek Laboratories, National University of
Singapore, 117542, Singapore
\\ $^{2}$Department of Physics, National University of Singapore, 117542, Singapore
\\$^{3}$Beijing-Hong Kong-Singapore Joint Center for Nonlinear and Complex Systems
(Singapore), National University of Singapore, 117542, Singapore
\\E-mail: tslwangj@nus.edu.sg and phylaich@nus.edu.sg}

\begin{abstract}

We study how to detect groups in a complex network each of which
consists of component nodes sharing a similar connection pattern.
Based on the mixture models and the exploratory analysis set up by
Newman and Leicht (Newman and Leicht 2007 {\it Proc. Natl. Acad.
Sci. USA} {\bf 104} 9564), we develop an algorithm that is
applicable to a network with any degree distribution. The
partition of a network suggested by this algorithm also applies to
its complementary network. In general, groups of similar
components are not necessarily identical with the communities in a
community network; thus partitioning a network into groups of
similar components provides additional information of the network
structure. The proposed algorithm can also be used for community
detection when the groups and the communities overlap. By
introducing a tunable parameter that controls the involved effects
of the heterogeneity, we can also investigate conveniently how the
group structure can be coupled with the heterogeneity
characteristics. In particular, an interesting example shows a
group partition can evolve into a community partition in some
situations when the involved heterogeneity effects are tuned. The
extension of this algorithm to weighted networks is discussed as
well.
\end{abstract}

\pacs{89.75.Hc, 89.75.Fb, 05.45.-a} \maketitle

\section{Introduction}
As a concise abstract model, the concept of network captures the
most essential ingredients of a complex system, namely, its basic
component units and their interaction configuration.  This
advantage --- simple in form but powerful in modelling --- has
attracted intensive studies of complex networks in a wide spectrum
of contexts, ranging from natural sciences to engineering problems
and human societies \cite{rev1,rev2,rev3}. Roughly speaking, the
investigations mainly fall into two categories: seeking the
topological characteristics and their origins in one and
understanding how they interact with the dynamical processes
supported by the networks in the other. It has been found that
topological characteristics, such as small-world \cite{sw} and
scale-free \cite{sf} properties, are quite general; they are
common features in a large set of networks from various fields.
Moreover, they are closely related to the dynamical processes on
the networks. Illuminating examples among many others include
epidemic spreading, to which the surprising implications of the
scale-free property have been well illustrated \cite{epid1,epid2};
and network synchronization, where the role played by the topology
can be marvellously separated and appreciated by analyzing the
master stability function \cite{sync}. Such progress has greatly
enhanced our belief in the significance of identification and
detection of these important topological characteristics
\cite{rev1,rev2,rev3}.

Community is another common topological feature that exists in
many complex networks.  Intuitively, a community refers to a set
of nodes whose connections between themselves are denser than
their connections to the nodes outside the set
\cite{comm1,comm2,comm3,comm4}. Community detection is very
important in network studies, because communities usually govern
certain functions as seen in many biochemical networks \cite{bio}
and social networks \cite{soci}. Communities also have important
implications to the dynamical processes based on the networks,
such as synchronization \cite{synrv1,synrv2,synrv3,synrv4},
percolation and diffusion \cite{diff1,diff2,diff3,diff4}. In
addition, in networks of large size, community structure may serve
as a crucial guide for reducing the network, which is believed to
be helpful in shedding light on the most essential properties of a
complex system \cite{redc1,redc2}. In view of the importance of
the community structure, there have been a lot of studies devoted
to the issue of community detection. (See Ref. \cite{cmrv} for a
recent and comprehensive review.) Recently, attempts have also
been made to extend the community detection methods developed in
these studies to weighted networks \cite{wtd1,wtd2} and directed
networks \cite{drct1,drct2}.

However, community is not the only perspective for partitioning a
network. For example, in a bipartite network, the best justified
partition is to separate all the nodes into two groups such that
nodes in one group only link to the nodes in the other. Indeed,
partition perspectives other than that of community is necessary
in order to have a better understanding of both the structures of
complex networks and the dynamical processes they support, as
shown in \cite{Ott} by the study of synchronous motions on
bipartite networks.

An insightful idea is to partition a network into groups where
nodes in each group share a similar connection pattern. As the
connection patterns are various and can vary from group to group,
this group model is very general and powerful in representing many
different types of structures in a network. This idea has a long
history. It was first introduced in social science by Lorrain and
White \cite{wt1}, where the nodes of similar connection pattern
are referred to as being {\it structurally equivalent}. This idea
has fruitfully led to the analysis of networks in social
\cite{wt2} and computer science based on block modelling. A recent
review can be found in Ref. \cite{gblk}.

In a recent study \cite{NL}, Newman and Leicht came up with a
novel and general partition scheme based on this idea. It divides
a network into groups of similar connection pattern. The most
striking advantage of their scheme lies in that it can be applied
for seeking a very broad range of types of structures in networks
without any prior knowledge of the structures to be detected. In
addition, the algorithm thus developed is ready to be used for
both the directed and undirected networks, and it is
straightforward to generalize it to analyze weighted networks
\cite{Ren}. The efficiency of the algorithm is also high in terms
of computation complexity. Recently, Ramasco and Mungan \cite{Ram}
have analyzed this method in detail and devised a generalized
Newman and Leicht algorithm based on their study. Other than the
Newman and Leicht algorithm and its variant \cite{Ram}, another
intriguing and insightful scheme for partitioning a network into
groups of similar connection pattern has also been developed based
on the information theory \cite{Rosv}.

The Newman and Leicht theory assumes that in a group the total
outgoing degree must be larger than zero \cite{Ram}. This
assumption limits the application of their theory. In order to
overcome this limitation, it has been suggested in \cite{Ram} to
deal with the incoming degrees, outgoing degrees, and
bidirectional degrees separately. In this paper, we show that by
assuming that all nodes in a group share the same {\it a prior}
probability to connect unidirectionally to a given node (see
analysis in Sec. III), this problem can be solved
straightforwardly. The algorithm we develop based on this
assumption can be applied without any restriction on the degree
distribution. Moreover, the partition of a network given by our
algorithm can be shown to be exactly the same as that of its
complementary network (see Sec. III). This is required by the
definition of a group of similar connection pattern. Another
advantage of our algorithm is that it allows an analysis of the
heterogeneity effects, which reveals further useful information of
the network structure. In addition to all of these, our algorithm
shows clearly that it is the {\it information} whether there is a
link between two given nodes, rather than the link exclusively (if
it exists between the two nodes), that contributes to the
partition. The information that there is no link between two given
nodes is $equally$ important. This insight provides a new and
different view for partitioning weighted networks. Our algorithm
also inherits all the advantages of that by Newman and Leicht.

In the next section, we first review briefly the theory by Newman
and Leicht, and then point out the extent of its applicability.
Next, in Sec. III, we develop our algorithm based on the {\it a
priori} probability assumption and discuss its properties. After
that we present examples of various types of groups together with
the analysis of two real networks. We discuss in Sec. IV the role
played by the involved heterogeneity effects, and show how a group
partition can depend on it by the example of the karate network
\cite{Zach}.  Finally, before summarizing the results of this
paper, we discuss in Sec. V how to extend our algorithm to
weighted networks.

\section{The Newman-Leicht algorithm (NLA)}

In search of the structures in a network, a dilemma we often
encounter is that we have to {\it input} initially what structures
we are intending to look for but this information is however
usually unavailable before the structures have been found
successfully. As a result what we can find eventually may strongly
depend on whether we have enough prior knowledge of the structures
to be detected. To overcome this difficulty, Newman and Leicht
\cite{NL} insightfully focused on the groups of similar connection
pattern. In their theory, the connection pattern for a group is
specified by sets of parameters to be determined. Initially, the
information of these connection patterns is not required as input
to the search algorithm thus designed; rather, they are shaped up
during the search process (running of the algorithm) and produced
as outputs. Finally, what the algorithm provides simultaneously is
not only the best way for grouping the nodes, but also the common
connection pattern that nodes in each group share. They made this
possible by skillfully harnessing the probabilistic mixture models
and the expectation-maximization algorithm \cite{NL}. As the
groups of similar connection pattern are effective in modelling
various structures in networks, their algorithm is very general
and has a wide application spectrum.

The main points of the Newman and Leicht theory are as follows.
(For the sake of convenience and clarity, we take the same
notation as in \cite{NL} throughout this paper.) Let us consider a
network of $n$ nodes belonging to $c$ groups. Its connection
configuration is given by the adjacency matrix $A$. If there is a
link between node $i$ and node $j$ then $A_{ij}=1$ otherwise
$A_{ij}=0$. In the Newman and Leicht theory, $n$, $c$ and $A$ are
assumed to be known and used as the input for their algorithm.
Here the number of groups $c$ is the only information needed in
advance about the partition. If it is unavailable, it should be
assumed or estimated based on other known information of the
network.

Next, the connection configuration $A$ is assumed to be a
realization of an underlying statistical model defined by two sets
of probabilities denoted by $\pi\equiv \{\pi_{r}\}$ and
$\theta\equiv \{\theta_{rj}\}$, respectively, with $r=1,\cdots,c$
and $j=1,\cdots,n$. This statistical model assumes that each node
has probability $\pi_{r}$ to fall in a group $r$ and for all nodes
in that group they have the same probability --- closely related
to $\theta_{rj}$ --- to connect to a given node $j$. Here
$\theta_{rj}$ is equivalent to the portion of the outgoing links
of group $r$ that connect to node $j$. The outgoing links of group
$r$ refers to the outgoing links that all nodes in group $r$ have.

In this sense $\theta_r \equiv \{\theta_{rj},j=1,\cdots,n\}$
defines the connection pattern shared by all nodes in group $r$.
As long as $\pi$ and $\theta$ are known, together with the
adjacency matrix $A$ as measured data, one can obtain the
probability for observing the node $i$ being in the group $r$,
namely $q_{ir}\equiv \Pr(g_i=r|A,\pi,\theta)$, and thus all the
information about the group partition. Here $g_i$ represents the
group to which the node $i$ is regarded to belong in a certain
partition; we use $q$ and $g$ to denote $\{q_{ir}\}$ and
$\{g_{i}\}$ respectively.

Hence the key is to specify $\pi$ and $\theta$. Newman and Leicht
assumed that the right values of the elements of $\pi$ and
$\theta$ are those that maximize the likelihood to observe the
connection configuration $A$ and a certain partition $g$, namely
$\Pr(A,g|\pi,\theta)$, or equivalently those that maximize its
logarithm
\begin{eqnarray}
L=\ln \Pr(A,g|\pi,\theta).
\label{eq21}
\end{eqnarray}
In this way, the problem is converted to a solvable fitting model
problem with the help of the maximum likelihood method \cite{NL}.
The next task is then reduced to find $\pi$ and $\theta$ that
satisfy this requirement.

To proceed further, Newman and Leicht adopted a crucial
simplification: they suggested instead to maximize the averaged
$L$ over all possible partitions:
\begin{eqnarray}
{\cal L}=\sum_{g_1}^c\cdots\sum_{g_n}^c \Pr(g|A,\pi, \theta) \ln
\Pr(A,g|\pi,\theta). \label{eq22}
\end{eqnarray}
As $\{g_i\}$ are summed out, this simplification allows one to
write down analytically the solutions of $\pi$ and $\theta$ in
terms of $A$ and $q$, and develop an efficient iterative algorithm
based on them. In detail, starting from
\begin{eqnarray}
\Pr(A|g,\pi,\theta)=\prod_{i,j} \theta_{g_i,j}^{A_{ij}}
\label{eq23}
\end{eqnarray}
and
\begin{eqnarray}
\Pr(g|\pi,\theta)=\prod_{i} \pi_{g_i},\label{eq24}
\end{eqnarray}
Newman and Leicht obtained
\begin{eqnarray}
\Pr(A,g|\pi,\theta)=
\prod_{i}\pi_{g_i}\prod_{j}\theta_{g_i,j}^{A_{ij}}\label{eq25}
\end{eqnarray}
and
\begin{eqnarray}
{\cal L} =\sum_{i,r} q_{ir} [\ln \pi_r+\sum_j
A_{ij}\ln\theta_{rj}]\label{eq26}
\end{eqnarray}
with
\begin{eqnarray}
q_{ir}=\frac{\pi_r \prod_j \theta_{rj}^{A_{ij}}}{\sum_s \pi_s
\prod_j \theta_{sj}^{A_{ij}}}.\label{eq27}
\end{eqnarray}
Then $\pi$ and $\theta$ that maximize $\cal L$ were deduced in
terms of $A$ and $q$ as
\begin{eqnarray}
\pi_r =\frac{1}{n} \sum_i q_{ir},\label{eq28}
\end{eqnarray}
\begin{eqnarray}
\theta_{rj}=\frac{\sum_i A_{ij} q_{ir}}{\sum_i k_{i} q_{ir}}
\label{eq29}
\end{eqnarray}
where $k_i\equiv \sum_j A_{ij}$ denotes the outgoing degree of
node $i$. Eqs. (\ref{eq27}), (\ref{eq28}) and (\ref{eq29}) thus
define the Newman-Leicht algorithm (NLA). It runs in an iterative
way: at each step, the old values of the elements of $q$, $\pi$
and $\theta$ are substituted into the right hand side of these
equations to generate their updated values. The convergent result
of $\theta$ then defines the connection patterns of groups and
that of $q$ suggests grouping. In practice, the calculation
converges rapidly. (We found that the convergence time goes as
$\sim O(n^2)$ in all the networks we have analyzed with the NLA,
including those that are not presented in this paper.)

\begin{figure}
\includegraphics[width=.9\columnwidth,clip]{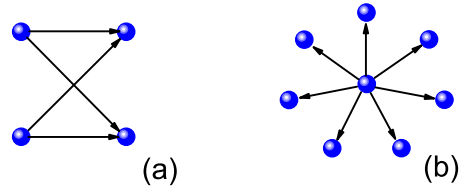}
\vspace{0.cm}\caption{Two examples where the Newman and Leicht
algorithm (NLA) does not apply. According to the definition, of
the two groups (of similar connection pattern) in the left network
(a) \cite{Ram} one contains the left two nodes and another
contains the right two; and of the two groups in the right network
(b) one consists of the center node and another consists of the
rest. However, due to the fact that one group, of the right two
nodes in (a) and the peripheral nodes in (b), has no outgoing
links, the Newman and Leicht algorithm (NLA) fails to partition
them correctly. As a comparison the APBEMA has no restriction on
the degree distribution; it partitions these two networks without
any ambiguity.} \label{fig1new}
\end{figure}

It should be noted that in getting Eqs. (\ref{eq28}) and
(\ref{eq29}) the following constraints imposed on $\pi$ and
$\theta$ have been taken into consideration:
\begin{eqnarray}
\sum_r \pi_{r}=1 \label{eq210}
\end{eqnarray}
and
\begin{eqnarray}
\sum_j \theta_{rj}=1 \label{eq211}. \label{eq211}
\end{eqnarray}
Indeed, the results given by Eqs. (\ref{eq28}) and (\ref{eq29})
satisfy these requirements. In addition, the results of Eq.
(\ref{eq28}) and Eq. (\ref{eq29}) are in consistency with the
definitions of $\pi_r$ and $\theta_{rj}$. In particular, Eq.
(\ref{eq29}) makes it clear that $\theta_{rj}$ is the expected
portion of the outgoing links of group $r$ that connect to node
$j$.

The definition of $\theta_{rj}$ and the corresponding
normalization condition imposed by Eq. (\ref{eq211}) imply that
the partition given by the NLA must be such that each group has at
least one outgoing link \cite{Ram}. This constraint limits the
application range of the NLA. An example cited in \cite{Ram} (see
Fig. 2 in \cite{Ram}) is a directed bipartite network which is
reproduced in Fig. 1(a). According to the definition of a group of
similar connection pattern, this network should be partitioned
into two groups such that one contains the left two nodes and one
contains the right two nodes, respectively. However, as the right
group has no outgoing links, NLA would suggest instead a partition
into the upper two nodes and the lower two nodes, or the whole
network as a single group \cite{Ram}. Another example is the
directed star as shown in Fig. 1(b); NLA partitions all nodes into
one group though from the viewpoint of similar connection pattern
or symmetry we expect the center node to be in one group and other
peripheral nodes in another.

\section{A priori probability based expectation maximization algorithm (APBEMA)}

In this section we present an expectation maximization algorithm
that does not have any restriction on the degree distribution of a
group. In addition, it also has many other advantages which will
be discussed in the following sections. Our method is in the same
spirit as the NLA, but the statistical model of the group is
different.

First let us suppose the network under consideration has $n$ nodes
that belong to $c$ groups, and the connection configuration is
given by the adjacency matrix $A$. Similarly, we assume $n$, $c$
and $A$ are known and serve as the input.

Next, as in the NLA, we assume that each node has probability
$\pi_r$ to fall in group $r$. $\pi_r$ in effect reflects the size
of group $r$, which is expected to be $n\pi_r$. As any node must
be in the network, we have
\begin{eqnarray}
\sum_r \pi_{r}=1. \label{eq31}
\end{eqnarray}

However, to specify the connection pattern of a group, we take the
{\it a priori} probability assumption instead. We assume that in a
given group $r$ all its nodes share the same {\it a priori}
probability, denoted by $\rho_{rj}$, to connect unidirectionally
to a given node $j$. As such $\rho_{rj}$ should satisfy $0\le
\rho_{rj}\le 1$. We also assume that $\rho_{ri}$ is independent of
$\rho_{rj}$ for $i\ne j$; namely, the probabilities for a node (in
group $r$) to connect to two different nodes are completely
independent. The normalization condition for $\rho_{rj}$ can be
expressed as $\rho_{rj}+(1-\rho_{rj})=1$, where $(1-\rho_{rj})$
stands for the probability with which a node in group $r$ does not
connect to node $j$. As compared with the NLA, here we need not
introduce a normalization condition like Eq. (\ref{eq211});
$\rho_{rj}$ can take any allowed value ($0\le \rho_{rj}\le 1$)
independently. It is this flexibility and adaptability that makes
our algorithm applicable in principle to any network.

Now we follow the NLA to develop the algorithm based on $\pi\equiv
\{\pi_{r}\}$ and $\rho\equiv \{\rho_{rj}\}$. In order to introduce
less notations, here we take all other symbols adopted in the NLA
except $\theta$ and maintain their original meaning (with $\theta$
being replaced by $\rho$ where necessary). We also refer to our
algorithm the {\it a priori} probability based expectation
maximization algorithm (APBEMA) in the following. Our starting
point is the conditional probabilities
\begin{eqnarray}
\Pr(g|\pi,\rho)=\prod_{i} \pi_{g_i} \label{eq32}
\end{eqnarray}
and
\begin{eqnarray}
\Pr(A|g,\pi,\rho)=\prod_{i,j} \rho_{g_i,j}^{A_{ij}}(
1-\rho_{g_i,j})^{1-A_{ij}}.
\label{eq33}
\end{eqnarray}
It should be stressed that the right hand side of Eq. (\ref{eq33})
accounts for not only the probability for the presence of a link
($A_{ij}=1$) but also that for a null link ($A_{ij}=0$), hence
honestly reflects the conditional probability for observing the
configuration given by $A$. As can be seen in the following, it
also implies the null links are as {\it equally} important as
links for partitioning a network, which agrees well with our
intuition.

Our next task is to find $\pi$ and $\rho$ that maximize
\begin{eqnarray}
{\cal L}=\sum_{g_1}^c\cdots\sum_{g_n}^c \Pr(g|A,\pi, \rho) \ln
\Pr(A,g|\pi,\rho). \label{eq34}
\end{eqnarray}
It can be rewritten as
\begin{eqnarray}
{\cal L}=\sum_{i,r}q_{ir}[\ln\pi_r+\sum_j
A_{ij}\ln\rho_{rj}\nonumber \\
 +\sum_j (1-A_{ij})(\ln(1-\rho_{rj})]
\label{eq35}
\end{eqnarray}
if we substitute Eqs. (\ref{eq32}) and (\ref{eq33}) into Eq.
(\ref{eq34}) with
\begin{eqnarray}
q_{ir} =\frac{\pi_r \prod_j
\rho_{rj}^{A_{ij}}(1-\rho_{rj})^{1-A_{ij}}}{\sum_s \pi_s \prod_j
\rho_{sj}^{A_{ij}}(1-\rho_{sj})^{1-A_{ij}}}.\label{eq36}
\end{eqnarray}
Here $q_{ir}\equiv \Pr(g_i=r|A,\pi,\rho)$. Apparently, it
satisfies the normalization condition $\sum_r q_{ir}=1$ as
required.

Now we are ready to obtain $\pi$ and $\rho$ that maximize $\cal L$
with the {\it only} constraint $\sum_r \pi_r=1$. We set
\begin{eqnarray}
f(\pi,\rho,\alpha)={\cal L} -\alpha(\sum_r \pi_r -1)\label{eq37}
\end{eqnarray}
with $\cal L$ being given by Eq. (\ref{eq35}) and $\alpha$ the
Lagrange multiplier introduced. By solving the following equations
\begin{eqnarray}
\frac{\partial f}{\partial \alpha}=0,~\frac{\partial f}{\partial
\pi_r}=0, ~{\text{and}}~ \frac{\partial f}{\partial \rho_{rj}}=0,
\label{eq38}
\end{eqnarray}
we obtain
\begin{eqnarray}
\pi_r =\frac{1}{n} \sum_i q_{ir}\label{eq39}
\end{eqnarray}
and
\begin{eqnarray}
\rho_{rj}=\frac{\sum_i A_{ij} q_{ir}}{\sum_i q_{ir}}.
\label{eq310}
\end{eqnarray}
Then we get the APBEMA defined by Eqs. (\ref{eq36}), (\ref{eq39})
and (\ref{eq310}). Its iterative implementation is the same as
that for the NLA, hence it has the same efficiency in terms of
computational complexity. Also as in the NLA, the convergent
values of $\{q_{ir}\}$ suggest the partition, and those of
$\{\rho_{rj}\}$ describe the connection patterns of groups.

It is worthwhile noting that according to Eq. (\ref{eq310}) $0\le
\rho_{rj}\le 1$ as expected. In addition, Eq. (\ref{eq310}) is
consistent with the meaning of $\rho_{rj}$, namely, the
probability with which a node in group $r$ is unidirectionally
linked to node $j$. This can be seen further from $\sum_j
\rho_{rj}$, which represents the averaged outgoing degree a node
in group $r$ has. Indeed, according to Eq. (\ref{eq310})
\begin{eqnarray}
\sum_j \rho_{rj}=\frac{\sum_i k_{i} q_{ir}}{\sum_i
q_{ir}}.\label{eq311}
\end{eqnarray}
($k_i\equiv \sum_j A_{ij}$ is the outgoing degree of node $i$.)
The right hand side of Eq. (\ref{eq311}) is exactly the expected
outgoing degree of a node in group $r$.

To summarize, our algorithm is based on the $a priori$ probability
assumption. It is this difference in the meaning between
$\rho_{rj}$ and $\theta_{rj}$ that makes the APBEMA radically
different from the NLA despite their similarity in form.

\subsection{Properties of the APBEMA}

The APBEMA developed previously has the following properties:

(i) {\it Applicable without any restriction on the degree
distribution}. Even in the trivial and less meaningful example
where the network contains some isolated nodes the APBEMA can
successfully assign them into one group, say group $r$, that is
characterized by $\rho_{rj}=0$. For the examples shown in Fig. 1,
the APBEMA partitions them without any ambiguity in the sense that
the output values of $\rho_{rj}$ and $q_{ir}$ are all virtually
zero or one. For the directed bipartite network shown in Fig. 1(a)
it suggests the left two nodes in one group and the right two in
another while for the directed star (Fig. 1(b)) it separates the
center node from the rest just as expected. (To apply the APBEMA
to these two networks, the number of groups has been assumed to be
$c=2$.)

(ii) {\it Suggesting the same partition for the complementary
network}. By the complementary network of a network specified by
the adjacency matrix $A$, we mean the network which has the same
nodes but its adjacency matrix $A'$ is related to $A$ via
$A'_{ij}=1-A_{ij}$. Namely, a link in network $A$ is a null link
in its complementary network $A'$ and vice versa. Obviously, a
group $r$ in $A$ characterized by $\{\rho_{rj}\}$ ($j=1,\cdots,n$)
is still a group in $A'$ with $\{\rho'_{rj}=1-\rho_{rj}\}$
according to the definition of group. Hence an algorithm aiming at
identifying the groups of similar connection pattern should
suggest the same partition for both a network and its
complementary network. This is the case for APBEMA, which is
guaranteed by the symmetry of $1-A_{ij}\to A'_{ij}$,
$1-\rho_{rj}\to \rho'_{rj}$, $\pi_{r}\to \pi'_{r}$ and $q_{ir}\to
{q'}_{ir}$ in Eqs. (\ref{eq36}), (\ref{eq39}) and (\ref{eq310}).
This symmetry also implies that null links play the same important
role as links in partitioning a network. A further discussion will
be given in Sec. V.

(iii) {\it Applicable to both directed and undirected networks}.
Although the APBEMA we obtain here is for directed networks, it
can be extended without any modifications in form to undirected
networks. The argument is similar to that given in \cite{NL}: In
an undirected network, $\rho_{rj}$ is still the probability for a
node in group $r$ to connect to node $j$; the probabilities for
there is and there is no link between node $i$ and node $j$ are
$\rho_{g_i,j}\rho_{g_j,i}$ and $(1-\rho_{g_i,j})(1-\rho_{g_j,i})$,
respectively. Hence
\begin{eqnarray}
&&\Pr(A|g,\pi,\rho)~~~~~~~~~~~~~~~~~~~~~~~~~~~~~~~~~~~~~~\nonumber\\
&&\quad=\prod_{i>j}\rho_{g_i,j}^{A_{ij}}\rho_{g_j,i}^{A_{ji}}(
1-\rho_{g_i,j})^{1-A_{ij}}(1-\rho_{g_j,i})^{1-A_{ji}}\nonumber\\
&&\quad=\prod_{i,j}\rho_{g_i,j}^{A_{ij}}(1-\rho_{g_i,j})^{1-A_{ij}},~~~
~~~~~~~~~~~~~~~~~~~~~ \label{eq312}
\end{eqnarray}
which is the same as Eq. (\ref{eq33}). ($A_{ij}=A_{ji}$ has been
used.) Other derivations are then exactly the same as in the
directed case.

(iv) {\it Powerful in accounting for the heterogeneity effects on
grouping.} The APBEMA allows us to prescribe the involved
heterogeneity effects of the outgoing degree distribution. This
can be done by conveniently introducing a tunable parameter to the
APBEMA. With this extension, we can study how the degree
heterogeneity may affect the grouping results in a controlled way.
In the situations where we desire to bias the heterogeneity
effects on the grouping this extended algorithm would be superior.
This algorithm will be discussed in detail in Sec. VI.


(v) {\it Applicable to weighted networks}. With a straightforward
extension, the APBEMA can also be used to analyze weighted
networks. A detailed discussion will be presented in Sec. V.

(vi) {\it The same efficiency as the NLA in terms of computational
complexity}.

\subsection{Examples}

To show how well the APBEMA works, we present in this subsection
several typical examples. Just as in the NLA, besides the
adjacency matrix $A$ we also need to set the number of groups,
$c$, as another input. For all the examples throughout this paper
we assume that this information has been known. In particular, we
set $c=2$ in all other examples except for the case of the
American college football teams where $c=12$ is assumed.

\begin{figure}
\includegraphics[width=.9\columnwidth,clip]{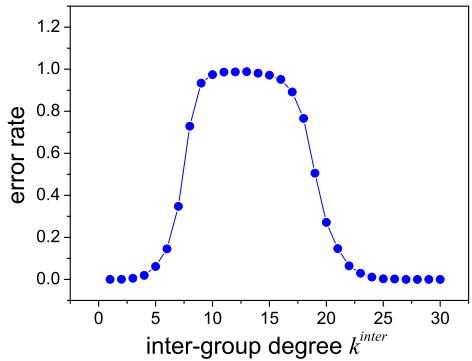}
\vspace{0.cm} \caption{An example for showing that the APBEMA can
identify the groups of similar connection pattern in a
$homogeneous$ network constructed according to the definition of
group. The network contains $n=60$ nodes which by construction are
divided into two sets of equal size. In each set the nodes are
randomly connected with the average intra-group degree
$k^{intra}=13$, and between the two sets the links are randomly
connected with the average inter-group degree $k^{inter}$. The
error rate by the APBEMA is shown as a function of the inter-group
degree $k^{inter}$. The two sets are successfully recognized for
$k^{intra}\gg k^{inter}$ and $k^{intra}\ll k^{inter}$ when the
group structure is clear.}
\end{figure}

The first example is a homogeneous undirected network. We simply
divide $n$ nodes into two sets of equal size and in each of them
nodes are randomly intra-connected with the average intra-degree
$k^{intra}$. After that the inter-group links are randomly added
with the average inter-group degree $k^{inter}$. Obviously, these
two sets are two groups according to the definition, and when
$k^{intra}\gg k^{inter}$ ($k^{intra}\ll k^{inter}$) they are
assortatively (disassortatively) connected. In practice, the
larger the difference between $k^{inter}$ and $k^{intra}$ is, the
clearer the group structure would be, and the easier it should be
to detect the groups.

The results for $n=60$, $k^{intra}=13$ against $k^{inter}$ are
summarized in Fig. 2.  We find that the APBEMA works well: it
identifies successfully both the assortatively and
disassortatively linked groups when their structures are clear. If
$k^{intra}$ and  $k^{inter}$ are too close it fails just as
expected.

It is interesting to note that when $k^{intra}\gg k^{inter}$ the
two groups can be seen as two communities. This fact suggests that
in the cases when groups and communities overlap with each other
in a network the APBEMA can be used to detect communities as well.
Given this, it is expected that for $k^{intra}\ll k^{inter}$, when
the network becomes bipartite-like, the APBEMA works equally well.
This is because the complementary network in this case is a
community network, and as having been pointed out in the last
subsection, the APBEMA is symmetric for a network and its
complementary network. Indeed, such a symmetry has manifested
itself clearly on the error rate curve presented in Fig. 2.

To measure the error of group detection, we define the error rate
$\epsilon$ as the sum of the portions of nodes wrongly partitioned
into the opposite group:
\begin{eqnarray}
\epsilon=\frac{\delta n_{12}}{n_{1}}+\frac{\delta n_{21}}{n_{2}}
\label{eq313}
\end{eqnarray}
where $n_{1}$ ($n_{2}$) is the number of nodes in the first
(second) group and $\delta n_{12}$ ($\delta n_{21}$) the number of
nodes belonging to group 1 (2) but are assigned to group 2 (1) by
the algorithm. If the nodes are randomly assigned to each group,
or all nodes are simply regarded as belonging to a single group,
the error rate so defined takes the value one and implies a
complete detection failure. It is zero only when all the nodes are
correctly grouped. To suppress the fluctuations, for every data
point presented in Fig. 2 we have averaged the error rates
evaluated over 1000 realizations of the network. We have also
checked that with other definitions of the detection error, for
example, that used in Ref. \cite{corl1,corl2,corl3}, which is
based on the normalized mutual information, the results are
qualitatively the same. This is also the case for all other
examples throughout this paper where the error rate is evaluated.

\begin{figure}
\includegraphics[width=.9\columnwidth,clip]{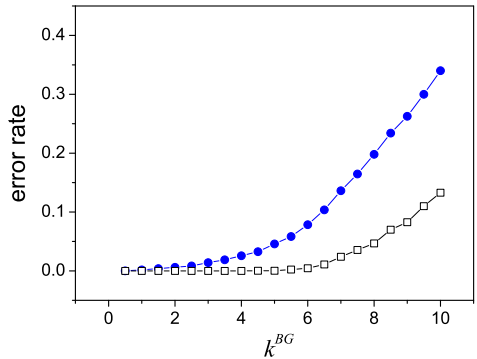}
\vspace{0.cm}\caption{An example for showing that the APBEMA can
identify the groups of similar connection pattern in a
$heterogeneous$ network constructed according to the definition of
group. The error rate (solid dots) is for the group detection
result by the APBEMA in identifying a fully connected clique of
$n_c=7$ nodes immersed in a randomly connected background of 63
nodes whose average degree $k^{BG}$ is varied for investigating
how the error rate depends on it. For $k^{BG}<n_c$ the APBEMA
works very well (the error rate is smaller than $<10\%$), and the
error rate due to wrongly partitioning the clique nodes into the
background (open squares) is small and can be neglected. In this
case the error rate is mainly contributed by wrongly partitioning
the background nodes into the clique as a result of fluctuations
in building the network. } \label{fig1}
\end{figure}

In our second example the groups are connected in a way neither
purely assortative nor purely disassortative. First we build a
random homogeneous and undirected network of $n$ nodes with the
average degree $k^{BG}$, then we chose from them $n_c\ll n$ nodes
randomly and fully connect them to form a clique. We then have two
sets of nodes: the clique, whose nodes have an average degree $
(n_c-1)+(1-n_c/n)k^{BG}$, and the one consists of the rest nodes
which we call the background, whose nodes have an average degree
$k^{BG}$. We restrict ourselves to the case $k^{BG}\ll n_c$,
namely, the degrees of the nodes in the clique are much larger
than those in the background, thus making the clique quite
outstanding to the background. Hence the network under
consideration is in fact highly heterogeneous. It should be
pointed out that in this case the communities occasionally formed
in the background due to fluctuations \cite{fluc} can be
neglected, and according to the definition the clique and the
background are two groups since nodes in themselves share the same
connection pattern that can be appropriately specified in terms of
$\{\rho_{rj}\}$. Furthermore, this network is neither assortative
nor disassortative; it is not a community network either because
the background nodes are connected between themselves the same
densely as they are connected to the clique nodes.

In Fig. 3 the partition results by the APBEMA for $n=70$ and
$n_c=7$ are shown against the average degree of the background
nodes, $k^{BG}$. It can be seen that for $k^{BG}\ll n_c$ it gives
the correct partition  perfectly. In fact, the APBEMA works well
all the way up to $k^{BG}\sim n_c$ with the error rate smaller
than 10\%. As $k^{BG}$ is increased further the clique becomes
less distinct from the background, and the fluctuations in the
background begin to play a role. As a result the error rate starts
to increase quickly. Further investigations show that for $k^{BG}<
n_c$ the detection error due to wrongly partitioning the clique
nodes into the background (open squares in Fig. 3), namely $\delta
n_{12}/n_{1}$ in Eq. (\ref{eq313})(subscript 1 (2) indicates the
clique (background)), is very small and can be safely neglected.
The detection error is mainly contributed by wrongly partitioning
the background nodes into the clique in certain network
realizations due to fluctuations where the wrongly partitioned
background nodes happen to have a higher degree and more links
connecting to the clique nodes. On average the total number of the
wrongly partitioned nodes (mainly from the background to the
clique) is about $0.11,0.39,0.88$ and $1.6$ for $k=1,2,3$ and $4$
respectively. In this calculation 1000 realizations of the network
are considered again to average the error rate.

\begin{figure}
\includegraphics[width=1.\columnwidth,clip]{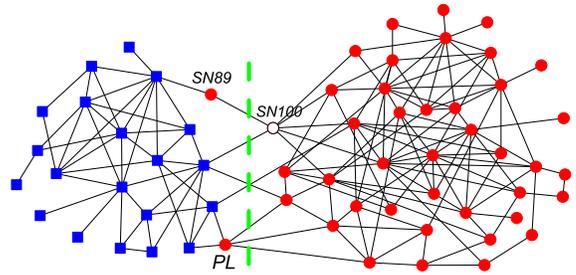}
\vspace{0.cm}\caption{The dolphin social network
\cite{dolf1,dolf2}. Nodes denoted by solid squares and solid dots
represent the two disjointed subdivisions the network split into
during the development of the network \cite{dolf3} after the
departure of a key member SN100 (open dot). The dashed line is the
group partition suggested by the APBEMA corresponding to the
largest value of $\cal L$ which regards nodes SN89 and PL
belonging to the opposite subdivision but all others nodes to
their own subdivisions. This is one real network example where the
APBEMA can be used to detect the community structure.}
\end{figure}

The network studied in this example could be relevant for studying
some real networks containing cliques. The success of the APBEMA
is a good indication of the flexibility and adaptability of the $a
priori$ probability assumption, and suggests that the APBEMA may
find some unique applications in certain partition problems.

In general, in a community network the nodes in a community may
not share the same connection pattern. In such cases the group
partition can be different from that of the community partition.
Such an example will be discussed in the next section. However, in
the cases where they do share the same connection pattern, or
approximately do, our algorithm can then be used to find the
community structure. This has been seen in the first example (Fig.
2) when the two groups are assortatively connected. In the
following we show two examples of real community network where the
partition result given by our algorithm is in good agreement with
the community partition.

The first one is a network of bottlenose dolphin \footnote{The
data of the network topology for the dolphin network and the
American football team network are downloaded from the website
http://www-personal.umich.edu/~mejn/netdata/.} living in Doubtful
Sound, New Zealand \cite{dolf1,dolf2,dolf3} which is composed of
62 dolphins (nodes) and 159 social ties (edges). It is assembled
by researchers over years (Fig. 4). During the course of the
investigation of this network, it split into two disjointed
subdivisions \cite{dolf3} of unequal size (represented by solid
squares and solid dots in Fig. 4 respectively) following the
departure of a key member named SN100 (denoted by the open dot in
Fig. 4). The group partition provided by the APBEMA corresponding
to the largest value of $\cal L$ agrees very well with the natural
splitting except two nodes named PL and SN89.

\begin{figure}
\includegraphics[width=1.\columnwidth,clip]{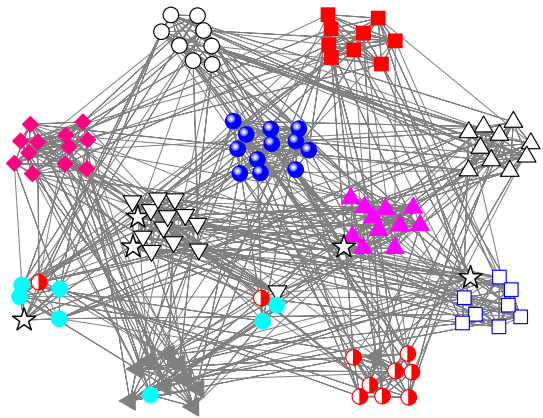}
\vspace{0.cm}\caption{The network of the American college football
teams extracted from the schedule of Division I games for the 2000
season \cite{foot}. The nodes denoted by the same symbols belong
to the same conference. The grouping result produced by the APBEMA
with assumed group number $c=12$ is represented by the clusters.
Stars stand for the ``IA independence'' conference which are
scattered due to their sparser connections inside. In this case
the groups given by the APBEMA coincide with the communities very
well despite the scattering of the ``IA independence'' conference.
This is another example in addition to the dolphin network (see
Fig. 4) where the APBEMA can be used to detect the community
structure. }
\end{figure}

The second example is the network of the American college football
teams \cite{foot}. The network is a map of the schedule of
Division I games for the 2000 season where 115 nodes represent the
teams and 616 edges represent regular-season games between the two
teams they connect \cite{foot}. All 115 teams are organized into
12 conferences each of which contains about 8-12 teams. As games
are usually more frequent between members of the same conference
than between members of different conferences, most conferences
can be seen as communities. But because there are few of them
whose teams played more or nearly as many games against teams in
other conferences than/as those in their own conference, the
network structure does not reflect the genuine conference
structure perfectly \cite{foot}.

The partition suggested by APBEMA is presented in Fig. 5. (The
number of the groups is assumed to be $c=12$ as input.) It can be
seen that the group structure suggested has a fairly accurate
coincidence with that of the conference. In particular, five
groups (the top five) are completely the same as the corresponding
conferences without any nodes wrongly assigned to/from other
conferences, and five others have only one or two nodes being
assigned to/from other conferences. The most obvious mismatch lies
in the partition of the conference ``IA independence''. Its
members, Central Florida, Connecticut, Navy, Notre Dame and Utah
State (denoted by stars in Fig. 5) are assigned to other groups
rather than in their own. Considering the fact that they have more
games in the conferences they are assigned to than in their own,
this is reasonable and somehow expected.

To summarize this subsection, the APBEMA performs well in
identifying various structures in a network. More examples and
further discussions of the presented ones will be given in the
following sections.

\section{Effects of heterogeneity on grouping}

In this section we study how the degree heterogeneity may affect
the grouping results. Theoretically this problem is interesting as
it is related to a general issue in network study, namely,
whether/how two different types of topological characteristics are
coupled. Obviously, in the APBEMA the coupling between the degree
distribution and the group structure is inherent: The APBEMA
suggests the grouping based on the connection patterns it
recognizes, but the connection patterns are in turn evaluated
based on the outgoing degrees. The close relation between the
connection patterns (given by $\{\rho_{rj}\}$) and the outgoing
degrees, $\{k_i\}$, can be seen clearly in Eq. (\ref{eq311}).

Then the next question for our aim here is how the APBEMA captures
the degree heterogeneity. A key observation is that the APBEMA
models the network in a coarse-graining way. It uses the groups as
the `patches' to represent different parts of the network, hence
in effect the network is characterized at two different levels. At
the lower level, namely inside each group, the APBEMA has assumed
that all nodes are identical and  statistically independent.
Therefore the structure of a group, its degree distribution as
well, has been assumed to be homogeneous. So at this level the
heterogeneity is not captured by the APBEMA, which can be seen as
a simplification adopted by the APBEMA. The difference between the
outgoing degree of a node from its expected value (i.e. $\sum_j
\rho_{rj}$, see Eq. (\ref{eq311})) in a group is treated by the
APBEMA as a result of the statistical fluctuations.

However, at the level of groups the APBEMA is flexible. It allows
the statistical characteristics of the groups to vary from group
to group so that the local structures of the network are given the
best matching. Therefore it is at this level that the
heterogeneity is taken into account by the APBEMA. With this
understanding we may imagine that the APBEMA tries to mimic the
degree distribution function with a series of peak-like functions.
Each peak-like function corresponds to a homogeneous degree
distribution in a group, and its position represents the average
outgoing degree of the group.

Hence if the network is heterogeneous, then the heterogeneity
would be characterized by the distances between these peaks. A
good example is the network studied in Fig. 3; its degree
distribution function happens to be one of two narrow peaks
representing the clique and the background. The distance between
them tells directly how heterogeneous the whole network is. For a
more general degree distribution function, though it is hard to
infer all the information of the heterogeneity based on the
distances between these peak-like functions, they are still a good
indicator of it. Another (opposite) extreme case is for the
homogeneous networks, see for example the one presented in Fig.2,
where all these peak-like functions overlap with each other and
the distances between them are all zero.

What we have learned here implies that if we can appropriately
preset the positions of these peak-like functions, namely the
average outgoing degrees of the groups, then we can interfere the
way the APBEMA considers the heterogeneity effects. Our aim in
this section is to develop such an algorithm. For example, if all
the average outgoing degrees are taken to be equal, then we have
in effect suppressed the heterogeneity effects to be considered
completely. This extreme case will be discussed in the first
subsection in the following. The APBEMA discussed in Sec. III  has
taken into account the heterogeneity effects as fully as it can,
so it stands as another extreme. In the second subsection we will
discuss how to introduce a control parameter to build an
interpolating algorithm such that the heterogeneity effects
involved can be tuned between these two extremes continuously.
Then we will show in the third subsection by the example of the
karate network \cite{Zach} how the heterogeneity plays its role in
grouping. A comparison with the dolphin network will reveal an
interesting underlying structural difference between the two
networks.

\subsection{The heterogeneity suppressed algorithm (HSA)}

As discussed in Sec. III, $\sum_j \rho_{rj}$ gives the expected
outgoing degree for a node in group $r$. If we assume that all the
nodes, regardless of which group they belong to, have the same
expected outgoing degree, then $\sum_j \rho_{rj}$ should satisfy
\begin{eqnarray}
\sum_j \rho_{rj}=\langle d^{out} \rangle, \label{eq41}
\end{eqnarray}
where $\langle d^{out} \rangle\equiv \frac{1}{n}\sum_{i,j}A_{ij}$
is the average outgoing degree over the whole network. With this
consideration, we can build up a grouping algorithm where the
effect of heterogeneity is completely suppressed. First we start
from Eqs. (\ref{eq32}) and (\ref{eq33}) and get $\cal L$ as in Eq.
(\ref{eq35}) and $q_{ir}$ as in Eq. (\ref{eq36}), namely,
\begin{eqnarray}
q_{ir}=\frac{\pi_r \prod_j
\rho_{rj}^{A_{ij}}(1-\rho_{rj})^{1-A_{ij}}}{\sum_s \pi_s \prod_j
\rho_{sj}^{A_{ij}}(1-\rho_{sj})^{1-A_{ij}}},\label{eq42}
\end{eqnarray}
again. Then we can get $\pi$ and $\rho$ with constraints of
$\sum_r \pi_r=1$ and those imposed by Eq. (\ref{eq41}) by setting
$f(\pi,\rho,\alpha,\beta)={\cal L}-\alpha(\sum_r \pi_r-1)-\sum_r
\beta_r(\sum_j \rho_{rj}-\langle d^{out}\rangle)$ and requiring
that the partial derivatives of $f$ with respect to its variables
to be zero. $\alpha$ and $\beta\equiv \{\beta_r\}$ serve as
Lagrange multipliers of the constrains. It leads to
\begin{eqnarray}
\pi_r =\frac{1}{n} \sum_i q_{ir},\label{eq43}
\end{eqnarray}
and
\begin{eqnarray}
\rho_{rj}=\frac{\beta_r \rho_{rj}^2 +\sum_i A_{ij}
q_{ir}}{\beta_r+\sum_i q_{ir}} \label{eq44}
\end{eqnarray}
with
\begin{eqnarray}
\beta_r=\frac{\langle d^{out} \rangle n\pi_r-\sum_i k_{i}
q_{ir}}{\sum_j \rho_{rj}^2-\langle d^{out} \rangle }. \label{eq45}
\end{eqnarray}
We refer to this algorithm defined by Eqs.
(\ref{eq42})-(\ref{eq45}) the heterogeneity suppressed algorithm
(HSA). As expected, if we impose zero to all $\beta_r$, then the
APBEMA is retrieved.

Compared with the APBEMA, the change in form of the HSA caused by
$\beta$ makes its implementation different: Here in fact two
cycles of iteration, the outer one and the inner one, are
involved. At each step of the outer cycle, we update $q$ and $\pi$
via Eqs. (\ref{eq42}) and (\ref{eq43}) first, then we come into
the inner cycle given by Eqs. (\ref{eq44}) and (\ref{eq45}) with
which the values of $\rho$ and $\beta$ are iterated till they
converge. Then a whole step of the outer cycle is finished. The
outer cycle is continued till all the values of $q$, $\pi$, $\rho$
and $\beta$ become stable. We notice that among various ways to
perform the inner iteration according to the equivalent transforms
of Eqs. (\ref{eq44}) and (\ref{eq45}) the one given by Eqs.
(\ref{eq44}) and (\ref{eq45}) is the best: It converges in all the
cases we have ever tested and the running time is the shortest.
(We find the running time also scales with $n$ as $\sim O(n^2)$
but is about two times of that consumed by the NLA and APBEMA.)

\subsection{The heterogeneity weighted algorithm (HWA)}

Now we have two extreme algorithms at hand: in one (the APBEMA)
the heterogeneity is given full consideration and in another (the
HSA) it is completely suppressed. Inspired by the way we construct
the HSA, we realize that an `interpolating' algorithm bridging the
two extremes can be created by introducing a tunable parameter $w$
into Eq. (\ref{eq41}) such that
\begin{eqnarray}
\xi_r(w)\equiv \sum_j \rho_{rj}=w\langle d_r^{out}
\rangle+(1-w)\langle d^{out} \rangle \label{eq46}
\end{eqnarray}
with
\begin{eqnarray}
\langle d_r^{out} \rangle\equiv\frac{\sum_i q_{ir}k_i}{\sum_i
q_{ir}}. \label{eq47}
\end{eqnarray}
Now $\xi_r(w)$ is the average outgoing degree we impose on the
group $r$, and the parameter $w$ prescribes the weight of the
heterogeneity. For $w=0$, $\xi_r(w=0)=\langle d^{out} \rangle$,
then no difference of the expected outgoing degrees between the
groups is considered; Eq. (\ref{eq46}) is then reduced to Eq.
(\ref{eq41}). For $w=1$, $\xi_r(w=1)=\langle d_r^{out} \rangle$,
which is exactly the average outgoing degree of group $r$ when the
heterogeneity is fully considered; it is then reduced to Eq.
(\ref{eq311}). For other values of $w$ ($0< w < 1$) the average
outgoing degree $\xi_r(w)$ takes the linear interpolating values
between $\xi_r(w=0)$ and $\xi_r(w=1)$ as a result.

Following the derivations as in the HSA, the solution of $\pi$ and
$\rho$ under constraints $\sum_r \pi_r=1$ and $\sum_j
\rho_{rj}=\xi_r(w)$ are still given by Eqs.
(\ref{eq42})-(\ref{eq44}), but $\beta_r$ now reads
\begin{eqnarray}
\beta_r=\frac{\xi_r(w)n\pi_r-\sum_i k_{i} q_{ir}}{\sum_j
\rho_{rj}^2-\xi_r(w)} \label{eq48}
\end{eqnarray}
instead. It is easy to show that for $w=0$ it reduces to Eq.
(\ref{eq45}) and the HSA is retrieved, and for $w=1$ as
$\beta_r=0$ we have the APBEMA again. For $0<w<1$ we thus have an
intermediate algorithm in between where only partial effects of
heterogeneity are considered, hence in effect it is a
heterogeneity weighted algorithm (HWA). By changing $w$ one can
therefore conveniently adjust the degree of heterogeneity involved
and investigate how it may affect the grouping results. The
numerical implementation of this algorithm is the same as the HSA.

As a trivial test this heterogeneity weighted algorithm has been
applied to the example in Fig. 2. As it is a homogeneous network,
we can expect that weighting the heterogeneity will not produce
any effects. Namely, the partition results shown in Fig. 2 does
not depend on $w$. Another trivial test is the clique-background
network studied in Fig. 3. As in this example the groups are
characterized by their own average degrees, we may expect that
suppressing the heterogeneity effects may blur the line of
distinction of the two groups and hence cause a detection
deterioration. These conjectures have been fully verified by our
simulations (the data of which are not shown here).

In the following we will consider some more meaningful and
inspiring examples. In particular we will apply the HWA to two
real social networks. Interesting results will be discussed in
detail.

\subsection{Analysis of the karate club}

In Ref. \cite{Zach}, Zachary reported an anthropological study of
a karate club in a university. During the development of the club,
two groups led by the instructor and the president formed
gradually and in the end, due to the lack of a solution to a
dispute, the club split.  In recent years, the network of this
karate club has been widely used for testing various community
finding techniques, including the NLA in \cite{NL} where it has
been found that the result of the NLA is in good agreement with
the true splitting.

\begin{figure}
\includegraphics[width=.7\columnwidth,clip]{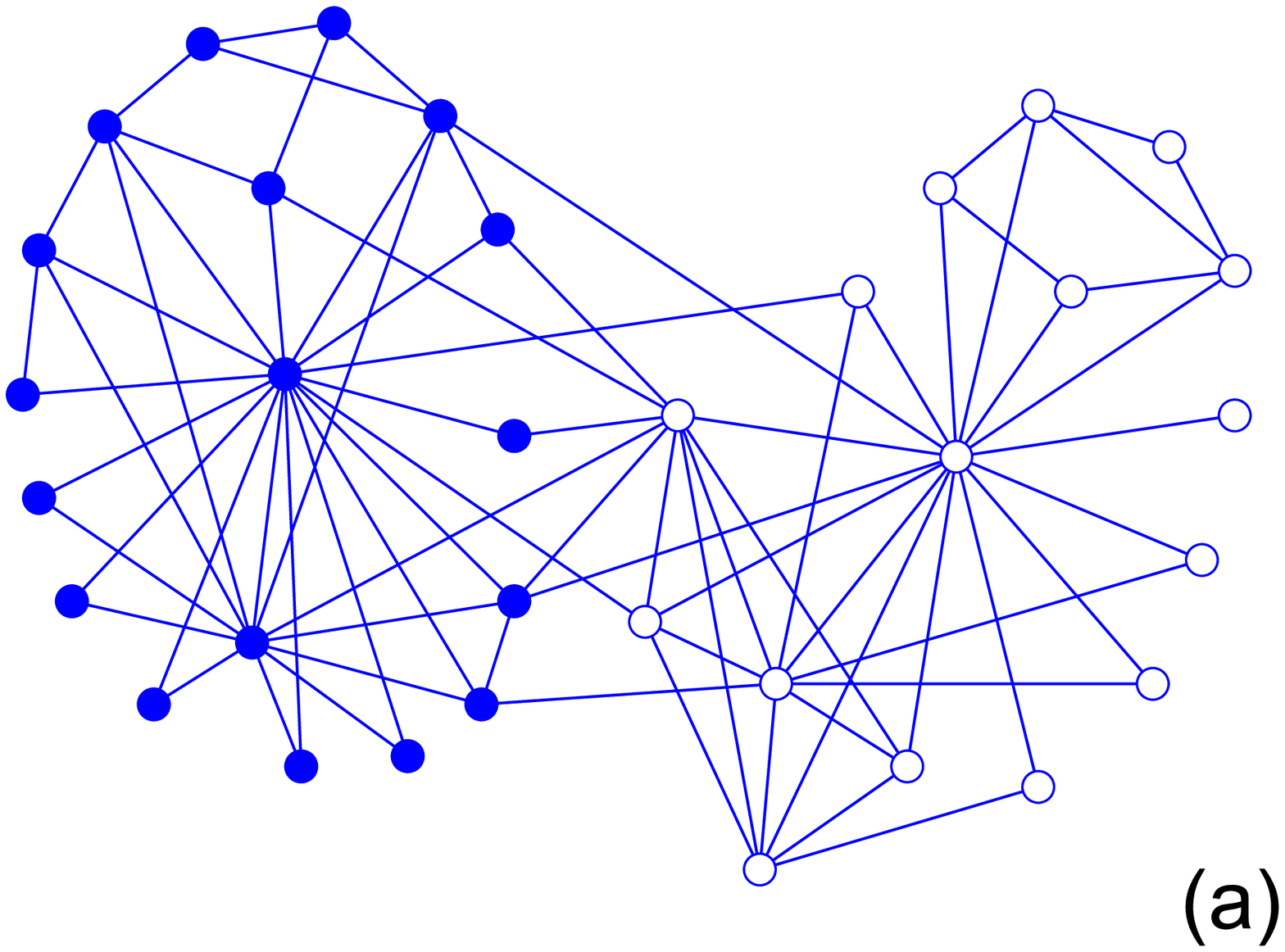}\vspace{-.8cm}
\includegraphics[width=.7\columnwidth,clip]{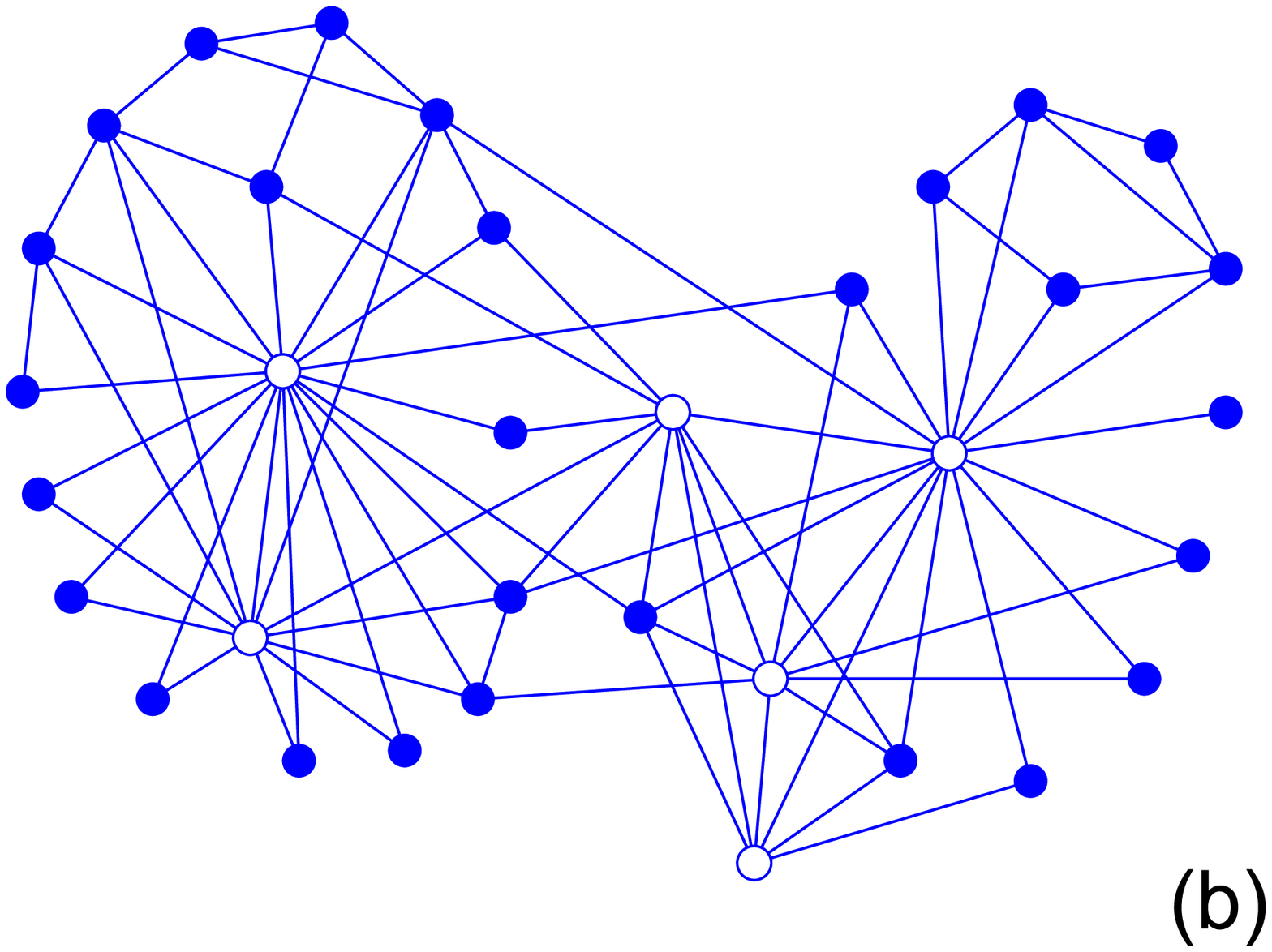}
\vspace{-.2cm} \caption{Grouping results for the karate club
network in a university \cite{Zach} given by the heterogeneity
suppressed algorithm (HSA) (a) and the APBEMA (b) respectively.
The two algorithms correspond to the special cases of $w=0$ (a)
and $w=1$ (b) of the heterogeneity weighted algorithm (HWA). The
groups are distinguished by different symbols representing the
nodes. The partition in (b) shows the groups may not be identical
with the communities in a community network.}
\end{figure}

To apply our heterogeneity weighted algorithm, it is found that
for $w=0$, namely the heterogeneity effects are completely
suppressed, the partition result is the same as that given by the
NLA (Fig. 6(a)). But for $w=1$ (Fig. 6(b)), when the heterogeneity
effects are fully considered, it suggests that those dominant
nodes (open dots in Fig. 6(b)) belong to one group and the others
belong to another group. Such a result (Fig. 6(b)) is not
surprising because nodes in each group are indeed much more
$similar$, which agrees better with our definition of group. For
example, nodes in each group have more similar degrees; they have
the similar connection pattern as well: in the dominant group
nodes are weakly connected to each other and serve as the branches
of the whole network, while in the other group nodes are only
sparsely connected between themselves and look like leaves
attached to the dominant group. This partition is also meaningful
in reality: it recognizes the leaders and coordinators from the
other members. It is important to note that from a different
viewpoint based on the information theory \cite{Rosv}, similar
partition result has been obtained (see Fig. 4B in \cite{Rosv}).
This example shows clearly that the groups of similar components
may not be the same as the communities in a community network.  In
order to have a better understanding of the network structure,
analysis of both is necessary.

Now let us look at what happens if the weight of the heterogeneity
is changed. Starting from $w=0$, each time we increase $w$ with a
small step $\Delta w$ and then iterate the stabilized results of
$q$, $\pi$ and $\rho$ obtained at $w$ until they converge. In this
way, we can trace the partition shown in Fig. 6(a) up to $w=1$.
Similarly, starting from $w=1$, the partition shown in Fig. 6(b)
can be traced back up to $w$ close to zero. The values of $\cal L$
evaluated by Eq. (\ref{eq35}) that correspond to these two
groupings are presented in Fig. 7. We can find that the
corresponding $\cal L$ value for the partition in Fig. 6(a)
changes only very slightly during this process, but that for the
partition in Fig. 6(b) is, first, smaller when $w$ is close to
zero, but it increases continuously with $w$ and at $w_c\approx
0.37$ it begins to become larger. For $w>w_c$, the fact that the
partition of Fig. 6(a) can still be traced suggests that the
corresponding value of $\cal L$ is, though not global, still a
local maximum as well. (As both partitions coexist for our
algorithm as maxima of $\cal L$, we believe that a network
analysis by the expectation maximization method would be more
powerful if local maxima solutions other than that of the global
maximum are considered in addition.)

\begin{figure}
\includegraphics[width=.9\columnwidth,clip]{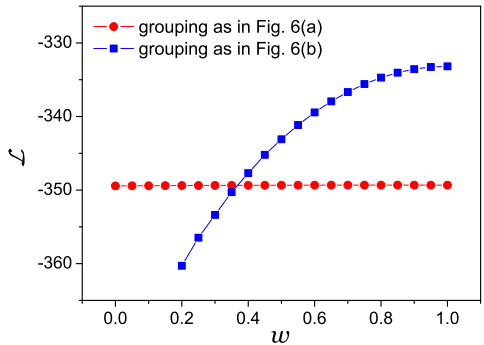}
\vspace{0.cm} \caption{Study on how the grouping of the karate
club network \cite{Zach} depends on the degree heterogeneity by
using the heterogeneity weighted algorithm (HWA). The $\cal L$
values corresponding to the two groupings shown in Fig. 6 are
presented as functions of $w$, the weight of the heterogeneity.
They are two maxima and intersect at $w_c\approx 0.37$. It
suggests that when the heterogeneity effects are suppressed
($w<w_c$) the partition as in Fig. 6(a) is preferred but when the
heterogeneity effects are more fully considered ($w>w_c$) the
partition as in Fig. 6(b) is recommended instead. It shows that a
group partition can depend on the heterogeneity effects strongly.}
\end{figure}

Fig. 7 shows clearly the important role played by the
heterogeneity in the definition and detection of the groups and
communities. In this example we have both groups and communities.
As they are identical for $w<w_c$, that is where our algorithm can
be used to detect the communities. If we insist that only the
solution corresponding to the global maximum of $\cal L$ defines
the groups, then they are different from the communities when
$w>w_c$.

On the other hand, as $w$ sets the weight of the heterogeneity to
be considered, this tunable algorithm is quite flexible and may
find some interesting applications in practice, in particular in
those situations where we wish to stress or weaken the effects of
the heterogeneity on purpose.

Next let us cite the social network of dolphin as a comparison. In
Fig. 8 the three largest maxima of $\cal L$ value are shown as
functions of the weight of the heterogeneity. There are not any
intersections between them. This fact may suggest that we have a
unique grouping and it is robust to the heterogeneity. This is
verified by the careful investigation that shows the partitions
corresponding to these curves indeed do not change with $w$. The
groupings corresponding to the largest two $\cal L$ maxima are
given by Fig. 4 and Fig. 9 respectively. A comparison between
these two partitions is interesting: the only difference lies in
the node PL. On one hand the nuance between their $\cal L$ values
may be a signature that our algorithm lacks confidence in
partitioning node PL due to its special role in between the two
subdivisions, and on the other hand their overwhelming agreement
may suggest that our algorithm is quite confident in partitioning
all other nodes except PL. This is consistent with the big gap
between the second and the third maxima of $\cal L$, which
indicates that our algorithm would prefer to discard any other
groupings except those shown in Fig. 4 and Fig. 9.

\begin{figure}
\includegraphics[width=.9\columnwidth,clip]{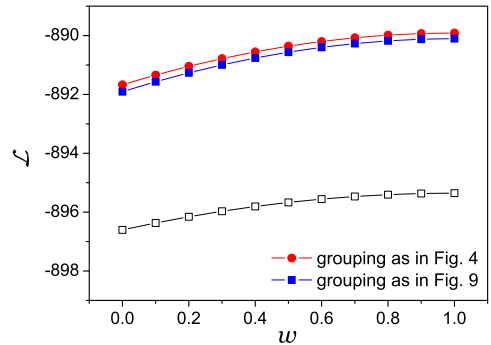}
\vspace{0.cm} \caption{Study on how the grouping of the dolphin
network \cite{dolf1,dolf2,dolf3} depends on the degree
heterogeneity by using the heterogeneity weighted algorithm (HWA).
The three largest maxima of the $\cal L$ value against the weight
of heterogeneity, $w$, are shown. The grouping of the network
corresponding to the top (middle) curve is given in Fig. 4 (Fig.
9). It suggests that in this example the group structure depends
insensitively on the heterogeneity effects.}
\end{figure}

These results may be an indication that the natural subdivisions
formed after the splitting of the network are the only main
topological structure from the view point of group partition in
this network. Unlike the karate network where different structures
may coexist, the network of dolphin lacks a `core' of dominant
nodes around which the other nodes are organized. This topological
difference may have implications in understanding the different
social behaviors of the two societies.

\section{Extension to the weighted networks}

As the expectation maximization algorithms have so many
advantages, it is desirable to extend them to weighted networks.
In fact the Newman and Leicht scheme favors such an extension. A
straightforward method was suggested in \cite{Ren} where the
weight of each link was related to its contribution to the $\cal
L$ value. In this section we discuss this problem based on the
APBEMA, but the derivations are similar and straightforward for
the heterogeneity suppressed and the heterogeneity weighted
algorithm. The radical difference between our scheme and that in
\cite{Ren} is that in our algorithm it is the {\it information}
provided by each entry of the adjacency matrix that is weighted.

\begin{figure}
\includegraphics[width=1.\columnwidth,clip]{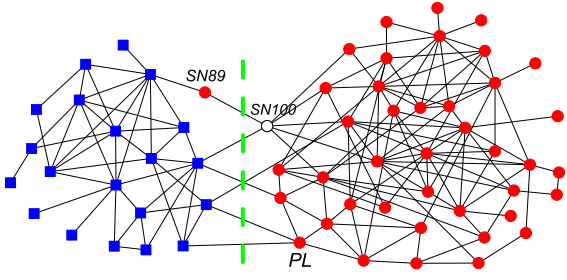}
\vspace{0.cm}\caption{The dolphin social network
\cite{dolf1,dolf2,dolf3}. Same as in Fig. 4 but the partition
represented by the dashed line, given by both the APBEMA and the
HWA, corresponds to the second maximum of $\cal L$ (see Fig. 8)
instead. In this partition only the node SN89 is not classified
into the natural subdivision it belongs to \cite{dolf3}. A
comparison with the partition corresponding to the first maximum
of $\cal L$ (see Fig. 4) indicates a special role node PL may
play.}
\end{figure}

We rewrite Eq. (\ref{eq35}) in the form of
\begin{eqnarray}
{\cal L} =\sum_{i,r} q_{ir} \ln \pi_r+\sum_{i,j}[
A_{ij}\sum_r q_{ir} \ln\rho_{rj}\nonumber\\
 + (1-A_{ij})\sum_r q_{ir} \ln(1-\rho_{rj})]\label{eq51}
\end{eqnarray}
from which we can tell that the term between the square brackets
represents the contribution to the $\cal L$ value given by
$A_{ij}$, namely the information of the connection state between
node $i$ and node $j$. Obviously, no matter $A_{ij}=1$ or
$A_{ij}=0$ its contribution is equally important and counts. Hence
if we attach a weight $\omega_{ij}$ to the information provided by
$A_{ij}$, then the $\cal L$ value for the aim of grouping should
naturally be replaced by
\begin{eqnarray}
\cal L^\omega&=&\sum_{i,r} q_{ir} \ln \pi_r+\sum_{i,j}\omega_{ij}[
A_{ij}\sum_r q_{ir} \ln\rho_{rj}\nonumber\\
 &&\qquad + (1-A_{ij})\sum_r q_{ir} \ln(1-\rho_{rj})]. \label{eq52}
\end{eqnarray}
Next, we assume the right grouping should be the one that maximize
$\cal L^\omega$ with the constrain $\sum_r \pi_r=1$. The deduction
is then the same as in the APBEMA and finally we have
\begin{eqnarray}
\pi_r=\frac{1}{n}\sum_i q_{ir}\label{eq53}
\end{eqnarray}
and
\begin{eqnarray}
\rho_{rj}=\frac{\sum_i q_{ir}\omega_{ij}A_{ij}}{\sum_i
q_{ir}\omega_{ij}},\label{eq54}
\end{eqnarray}
where $q_{ir}$ is still given by Eq. (\ref{eq36}). It is apparent
that, for an unweighted network where $\omega_{ij}={\text
{const}}$, this algorithm is reduced to the APBEMA as expected.

Similarly, if the constraints of Eq. (\ref{eq41}) or Eq.
(\ref{eq46}) are taken into account, we can get the heterogeneity
suppressed or heterogeneity weighted algorithm for the weighted
network as well.

It is important to note that $\omega_{ij}$ is the weight of the
information provided by $A_{ij}$ rather than of the link between
node $i$ and $j$. (Note that though in calculating $\rho_{rj}$
(Eq. (\ref{eq54})) $\omega_{ij}$ does not count in evaluating the
numerator if $A_{ij}=0$, it does in evaluating the denominator.)
In other words, even if there is no link between node $i$ and node
$j$, this piece of information ($A_{ij}=0$) is equally important
for recognizing the group structure. This result is consistent
with our intuition and experience.

In order to well appreciate the implications of this algorithm,
let us take the network studied in Fig. 3 as an illustration. For
the sake of simplicity, we assume that all the weights take only
two values: 1 and $\omega$. Here $\omega$ is a constant used to
weight a selected potion of entries of the adjacency matrix and
$0\le \omega\le 1$; it is introduced to control the information of
that potion the algorithm can use and so that we can investigate
how the grouping results depend on it. We consider the following
three cases: (i) $\omega_{ij}=1$ for $A_{ij}=0$ and
$\omega_{ij}=\omega$ for $A_{ij}=1$; (ii) $\omega_{ij}=1$ for
$A_{ij}=1$ and $\omega_{ij}=\omega$ for $A_{ij}=0$; (iii)
$\omega_{ij}=\omega$ if both node $i$ and node $j$ are in the
clique and $\omega_{ij}=1$ otherwise. For $\omega=0$, since a
crucial part of information of the network topology lacks, we may
expect a failure of grouping. As $\omega$ is increased, more and
more information are taken into account, the grouping should be
more and more accurate. Finally, as $\omega=1$ is approached, all
the topological information is considered, our algorithm should
suggest the grouping as perfectly as the APBEMA does. This
conjecture has been well verified by the simulations. In Fig. 10
the grouping error rate against $\omega$ is summarized for the
case where the network has $n=70$ nodes, the clique size is
$n_c=7$ and the average degree of the background nodes $k^{BG}=3$.
Each data point represents the averaged error rate over 1000
realizations of the network.

In the first case (solid squares in Fig. 10), the information
associated with the null links is fully considered but that
associated with the links is controlled by $\omega$. For
$\omega=0$ their contributions are completely ignored; as a
consequence the algorithm `sees' all the nodes isolated from each
other and classifies them into a single group. To increase
$\omega$ from zero, thought slightly, would stop the algorithm
from classifying all the nodes in a single group, but the error
rate is still high. As $\omega$ is increased further, more and
more information of the links is available and the partition
becomes more and more accurate. When it comes to the point
$\omega\sim 0.7$, the information seems to have been enough for
the algorithm to recognize well the clique from the background.
This phenomenon is interesting: it suggests that in fact there is
a redundance in information for the use of partition in the
network under study.

In the second case (open squares in Fig. 10), the information
associated with the links is fully considered but that with the
null links is tuned by $\omega$. Similarly, for $\omega=0$ the
algorithm cannot `see' the null links and thus the background. All
nodes are regarded to be in one well connected group. This result
shows clearly the information of the null links is a requisite for
a correct partition. As $\omega$ is increased from zero, the error
rate undergoes an abrupt drop. This is because here we have much
more null links than links and hence even a small value of
$\omega$ may release much more information than in the case (i).
To increase $\omega$ further would improve the grouping
correspondingly just as expected.

In the last case (solid dots) the weights of the information
associated with the clique is varied instead, but again we have
qualitatively the same result as in the first two cases. These
results are in good consistency with our discussions on the
weighted APBEMA from the information perspective.

\begin{figure}
\includegraphics[width=.9\columnwidth,clip]{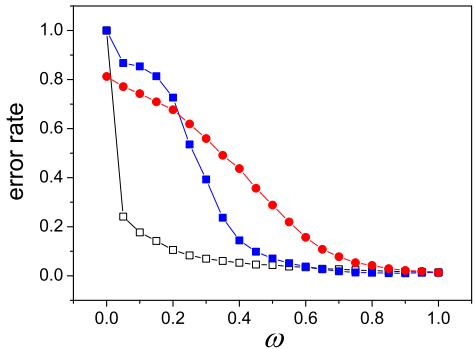}
\vspace{0.cm}\caption{Error rates for the grouping results
suggested by the weighted APBEMA in identifying a fully connected
clique of $n_c=7$ nodes immersed in a randomly connected
background of 63 nodes whose average degree is $k^{BG}=3$. The
information contained in each entry of the adjacency matrix is
weighted by either $\omega$ or 1, and solid squares, open squares
and solid dots represent three different ways for assigning the
weights among the entries, which correspond to the case (i), (ii)
and (iii) as described in the text (see the text). In all the
cases as $\omega$ is increased the grouping becomes more accurate,
which supports the viewpoint that the information of links and
null links are equally important.}
\end{figure}

To weight the information contained in $\{A_{ij}\}$ can be more
relevant in practice. To construct a network representation of a
real complex system, it involves unavoidably the measurement of
the connection state between any two nodes. In a general case, the
measurement does not generate a definite zero/one output; rather,
the errors and uncertainties are entangled intrinsically. In many
cases, such as in some biological systems, biochemical systems and
human societies, as the relations between the elements can be
numerous and of various types on one hand, and these relations
themselves can be coupled with each other on the other hand, the
problem of measurement is even more subtle and difficult. Hence
for any network abstracted in the end, the evaluations of the
confidence in the measured connection states are important and
necessary. These evaluations of the confidence are the ideal
measures of the weights considered here.

\section{Summary}

In this work we have studied how to detect the groups in a complex
network that consist of nodes having the similar connection
pattern. Our algorithm is based on the mixture models and the
exploratory analysis suggested by Newman and Leicht, but
significant differences exist. In our algorithm the connection
pattern is modelled by the {\it a priori} probability assumption
instead. The main advantages of our algorithm are that (i) It can
be applied without any restriction on the degree distribution;
(ii) It possesses the symmetry between the links and the null
links; (iii) It is flexible in dealing with the heterogeneity
effects; and (iv) It can be extended to the connection information
weighted networks. These advantages have been illustrated by
various network examples.

With our algorithm we have studied the role played by the
heterogeneity. We find that the grouping result may depend on the
heterogeneity effects involved. This finding suggests that in
order to have a thorough knowledge of the network structure, this
dependence should be analyzed. For this reason all the groupings
found (at various values of $w$, see Sec. IV) are justified. This
can be seen as an extension to the definition of group formally
defined at $w=1$ when the heterogeneity effects are fully
considered.

Based on our analysis, it is natural to extend our algorithm to
the connection information weighted networks. This result is a
direct implication of our {\it a priori} probability based group
connection pattern model. As the connection information weighted
networks can be closely related to the measurement of networks, we
expect our extended algorithm may find wide applications.

Finally, our study has also suggested that groupings associated
with other top maxima of the merit function ($\cal L$) could be
meaningful and useful as well. This may be a common feature among
the expectation maximization algorithms. How to interpret these
groupings seems to be interesting and potentially important that
deserves further investigations.

\acknowledgments

This work is supported by Defense Science and Technology Agency
(DSTA) of Singapore under agreement of POD0613356.

\end{document}